\documentclass[twocolumn]{aastex61}

\usepackage{amsmath}
\usepackage{bm}
\usepackage{graphicx}

\newcommand{\deriv}[2]{\frac{d{#1}}{d{#2}}}

\newcommand{\psrB}{B0031$-$07}
\newcommand{\phase}{\varphi}

\begin{document}

\title{The frequency-dependent behaviour of subpulse drifting: I. Carousel geometry and emission heights of PSR \psrB{}}

\author{S. J. McSweeney}
\author{N. D. R. Bhat}
\affiliation{International Centre for Radio Astronomy Research (ICRAR), Curtin University, 1 Turner Ave., Technology Park, Bentley, 6102, W.A., Australia}
\author{G. Wright}
\affiliation{Jodrell Bank Centre for Astrophysics, School of Physics and Astronomy, University of Manchester, M13 9PL, UK}
\author{S. E. Tremblay}
\affiliation{International Centre for Radio Astronomy Research (ICRAR), Curtin University, 1 Turner Ave., Technology Park, Bentley, 6102, W.A., Australia}
\author{S. Kudale}
\affiliation{National Centre for Radio Astrophysics, Tata Institute of Fundamental Research, Pune 411007, India}

\begin{abstract}

The carousel model of pulsar emission attributes the phenomenon of subpulse drifting to a set of discrete sparks located very near the stellar surface rotating around the magnetic axis.
Here, we investigate the subpulse drifting behaviour of PSR \psrB{} in the context of the carousel model.
We show that \psrB{}'s three drift modes (A, B, and C) can be understood in terms of a single carousel rotation rate if the number of sparks is allowed to change by an integral number, and where the different drift rates are due to (first-order) aliasing effects.
This also results in harmonically-related values for $P_3$ (the time it takes a subpulse to reappear at the same pulse phase), which we confirm for \psrB{}.
A representative solution has $[n_A,n_B,n_C]=[15,14,13]$ sparks and a carousel rotation period of $P_4 = 16.4\,P_1$.
We also investigate the frequency dependence of \psrB{}'s subpulse behaviour.
We extend the carousel model to include the dual effects of aberration and retardation, including the time it takes the information about the surface spark configuration to travel from the surface up to the emission point.
Assuming these effects dominate at \psrB{}'s emission heights, we derive conservative emission height differences of $\lesssim 2000\,$km for mode A and $\lesssim 1000\,$km for modes B and C as seen between $185\,$MHz and $610\,$MHz.
This new method of measuring emission heights is independent of others that involve average profile components or the polarisation position angle curve, and thus provides a potentially strong test of the carousel model.

\end{abstract}

\section{Introduction}

Subpulse drifting, first observed by \citet{Drake1968} not long after pulsars were discovered \citep{Hewish1968} is a widespread phenomenon affecting more than half of all known pulsars \citep{Weltevrede2006,Weltevrede2007}.
When individual pulses are stacked vertically to form a two dimensional pulsestack, the individual components of pulses, termed \textit{subpulses}, are observed to undergo a regular modulation pattern in both amplitude and phase, which visually resemble a set of discrete diagonally-oriented burst regions called \textit{drift bands}.
The slope of the drift bands is called the drift rate, and it represents the apparent advance or lag of subpulses per stellar rotation.

Drifting subpulses found an early explanation in the seminal polar gap theory of \citet{Ruderman1975} (hereafter RS), who suggested that the pattern of radio waves that ultimately escape the magnetosphere is the emission signature of a set of discrete, localised pockets of quasi-stable electrical activity called \textit{sparks} that exist very near the pulsar surface.
These sparks are the sites of particles accelerated through the polar gap to energies of $10^{12}\,$V, seeding an avalanche of secondary particles which stream along magnetic field lines at relativistic speeds to produce curvature radiation.
 Charged particles generated by the sparks partly screen the electric potential along the magnetic field at the surface, causing them to move relative to the polar cap surface about the magnetic axis (the $\mathbf{E}\times\mathbf{B}$ drift), in an arrangement resembling a fairground ``carousel''.
The emission beam would then also have a spatial structure composed of discrete ``beamlets'' that reflects the magnetic azimuthal arrangement of sparks on the surface.
With each rotation, an Earth-bound observer perceives a different intensity pattern as the line of sight cuts through a slightly rotated carousel, producing the observed drifting behaviour.

The carousel model has been heavily criticised on (at least) two different grounds.
Firstly, its reliance on a coherent form of curvature radiation is difficult to justify from known plasma physics \citep[e.g.][]{Melrose1992}.
Secondly, there are a number of secondary phenomena associated with subpulse drifting that present challenges for, or at least require extensions of, the simple geometric picture described above \citep{Edwards2006}.
These include, among others, drift mode switching \citep{Huguenin1970}, bi-drifting \citep{Qiao2004}, nulling \citep{Backer1970b}, and subpulse ``memory'' across nulls \citep{Lyne1983}.

Despite these criticisms, the carousel model, along with its implied conal emission beams, remains a popular model for interpretating subpulse drifting, as it is capable of explaining (qualitatively at least) several features of pulsar morphology and behaviour \citep[][and subsequent papers in the series]{Rankin1983}.
As a result, there have arisen two schools of thought, one which believes the basic model is correct in principle but whose details are complicated and have yet to be worked out on a pulsar-by-pulsar basis, and one which believes the model is fundamentally incorrect.
What is lacking is a robust test that can distinguish, once and for all, between these alternatives.
A strong candidate for such a test, if it can be found, is one that exploits the model's unique emission geometry, namely, that although the observed photons are emitted at some height dictated by the magnetic field geometry, the subpulse modulation pattern is dictated by events very close to the surface.
This feature of the carousel model makes specific predictions about how the observed subpulse modulation changes as a function of observing frequency.

In this two-part paper series that will focus on \psrB{}, we investigate the frequency dependence of subpulse drifting.
We start by recognising that the basic carousel model predicts that the rotation phases at which subpulses appear is invariant with frequency \citep{Edwards2002,Edwards2003}, a point that is sometimes confused with the expected radius-to-frequency mapping (RFM) behaviour of average profiles \citep[e.g.][]{Smits2007,Yuen2016}.
Nevertheless, frequency-dependent behaviour of subpulses \emph{is} observed, most notably in the way that the time between successive subpulses (the so-called ``secondary period'', $P_2$) changes as a function of frequency \citep{Taylor1975,Bartel1980}.
Several ideas have been advanced to explain this within the context of the carousel model, including non-dipolar field geometries \citep[e.g.][]{Davies1984} and the finite size and shape of beamlets, which present a slightly-shifted cross section to the line of sight as the frequency-dependent beamlets move in magnetic colatitude \citep{Edwards2003,Bilous2018}.
\citet{Edwards2003} also speculate that aberration and retardation (AR) effects must be present, although they argue that the subpulse phase shifts they observe in PSR B0320+39 and PSR B0809+74 imply emission heights that are much larger than the heights expected from other arguments.

In this paper, we develop this idea further to obtain quantitative predictions of how subpulses shift with frequency if AR effects are the dominant cause.
We apply the model to the interesting and well-studied case of PSR \psrB{} whose $P_2$ is also known to have a measurable longitude dependence \citep{McSweeney2017} and whose known frequency-dependent subpulse phase shifts are also dependent on which of the pulsar's three drift modes is present \citep{Huguenin1970,Vivekanand1996,McSweeney2017a}.
In Sections \S\ref{sec:drift_constraint} and \S\ref{sec:P2_constraint} the relationship (initially neglecting AR effects) between the drift bands and the underlying carousel geometry are laid out, with particular emphasis on the degeneracy introduced by the possibility of aliasing effects.
We then generalise the model to include AR effects in Section \S\ref{sec:ar}, calculating their effect on the subpulse phase shift (\S\ref{sec:ar_sps}) and on $P_2$ (\S\ref{sec:ar_P2}).
A discussion of the results and our final conclusions are presented in Sections \S\ref{sec:discussion} and \S\ref{sec:conclusions}.

This is the first of a two-part series, both of which deal with the relationship between frequency-dependent effects and subpulse drifting in \psrB{}.
This paper deals exclusively with effects that can be observed in Stokes I observations.
In the subsequent paper, we investigate how the subpulse phase (which encodes information about the magnetic azimuth) is related to the polarisation position angle and use this relationship to place further constraints on the carousel and viewing geometries.

\subsection{Definitions and Notation}
\label{sec:def}

\begin{figure}
    \centering
    \includegraphics[scale=0.3]{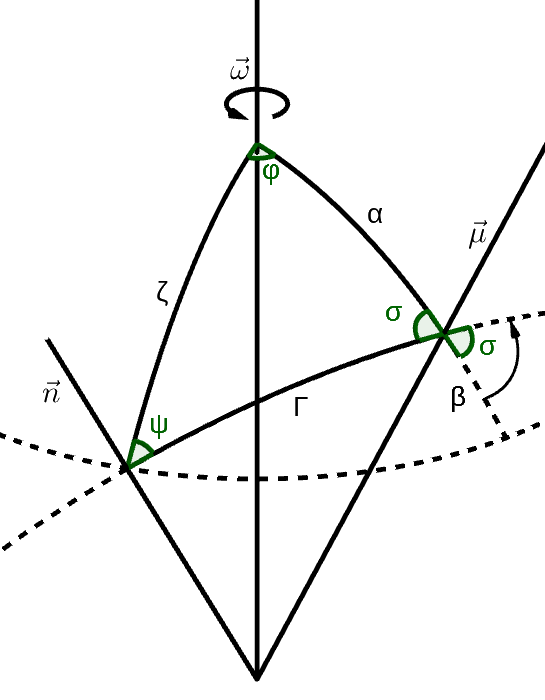}
    \caption{The fundamental spherical triangle of pulsar viewing geometry. The LoS has been labeled $\vec{n}$. The other symbols are defined in the accompanying text.}
    \label{fig:pulsarangles}
\end{figure}

The rotation axis ($\vec{\omega}$), the magnetic axis ($\vec{\mu}$), and the LoS define three points of a spherical triangle which relate the rotation phase ($\phase$), the viewing geometry ($\alpha$, $\zeta$), the half-beam opening angle ($\Gamma$), the PPA ($\psi$), and the magnetic azimuth ($\sigma$).
In this paper, $\alpha$ and $\zeta$ are always relative to the rotation axis defined by the right hand rule, so that $0^\circ \le \alpha < 180^\circ$ and $0^\circ \le \zeta < 180^\circ$.
These angles are illustrated in Fig. \ref{fig:pulsarangles}.
An asterisk is used to denote the supplementary angles $\alpha^\ast = 180^\circ - \alpha$ and $\zeta^\ast = 180^\circ - \zeta$, i.e. measured from the opposite rotation axis.
The impact angle is defined as $\beta = \zeta - \alpha$, with $\beta^\ast = \zeta^\ast - \alpha^\ast = -\beta$.

As is common in the literature, we use $P_1$ for rotation period of the pulsar, $P_2$ for the subpulse separation, $P_3$ for the time it takes a subpulse to reappear at the same rotation longitude, and $P_4$ for the rotation period of the carousel.
Conventionally, $P_2$ is often expressed in degrees of rotation, and $P_3$ in units of $P_1$, but in order to keep the notation consistent throughout this work, we reserve the unadorned $P_\text{x}$ to have units of time.
To denote quantities normalised by $P_1$, we introduce the overline notation, e.g., $\overline{P}_3 \equiv P_3/P_1$.
In addition, we define $P_2^{(\circ)} \equiv \overline{P}_2 \times 360^\circ$ to express periods in degrees of rotation.

The drift rate, $D$, is traditionally expressed in units of degrees per pulse.
To keep this consistent with the foregoing notation, we define
\begin{equation}
    D \equiv \frac{P_2}{\overline{P}_3}, \quad
    \overline{D} \equiv \frac{\overline{P}_2}{\overline{P}_3} \quad
    \text{and} \quad
    D^{(\circ)} \equiv \frac{P_2^{(\circ)}}{\overline{P}_3}.
\end{equation}

\section{Observations}
\label{sec:obs}

We have taken two simultaneous observations of \psrB{}, one using the Murchison Widefield Array \citep[MWA;][]{Tingay2013} centred at $185\,$MHz, and the other using the Giant Metrewave Radio Telescope \citep[GMRT;][]{Swarup1991}, centred at $610\,$MHz.
The details are summarised in Table \ref{tbl:observations}.
\begin{deluxetable}{ccccc}
    \tablecaption{Data sets recorded on MJD 57556\label{tbl:observations}}
    \tablehead{
        \colhead{Telescope} & \colhead{Centre frequency} & \colhead{Bandwidth} & \colhead{\# pulses} & \colhead{S/N per} \\
        & \colhead{(MHz)} & \colhead{(MHz)} & & \colhead{pulse}
    }
    \startdata
        MWA  & 185 & 30.72 & 4631 & 14.2 \\
        GMRT & 610 & 33.33 & 8487 & 9.8 \\
    \enddata
\end{deluxetable}

The MWA observation was processed in the same way as \citet{McSweeney2017}.
The MWA voltages were recorded using the Voltage Capture System \citep[VCS;][]{Tremblay2015}, and a tied-array beam was formed \citep{Ord2019} in the direction of the pulsar.
The calibration solution was obtained using a separate, dedicated observation of the calibrator source 3C444.
During the calibration process, it was deemed necessary to exclude five MWA tiles (out of 128) from the coherent sum, due to corrupted data.
The resulting data set had a time resolution of $100\,\mu$s and a frequency resolution of $10\,$kHz.
At the time of the observation (June 2016), only total intensity data products were available, but the recent verification of MWA polarimetry \citep{Xue2019} has made it possible to reprocess the original recorded voltages to produce a data product with full Stokes parameters.

The GMRT observations were made with the 13 central antennas in the phased array mode.
However, only the total intensity was recorded.
The output time resolution was $123\,\mu$s and the frequency resolution was $65\,$kHz.

\begin{figure}
    \centering
    \includegraphics[scale=0.35]{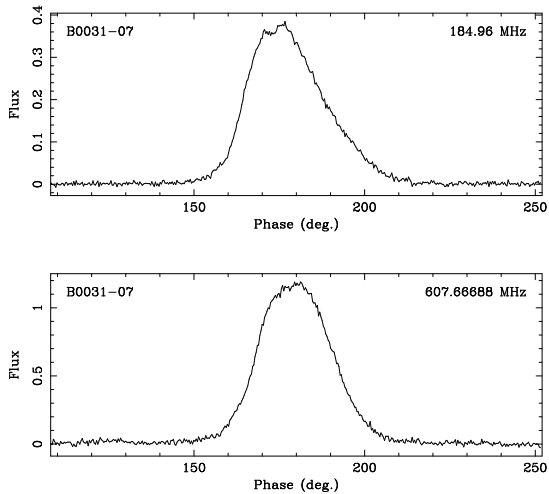}
    \caption{Average intensity profiles of \psrB{} made from the MWA observation (upper panel) and the GMRT observation (lower panel). The alignment of the profile centres was done by eye.}
    \label{fig:obs3_psrchive}
\end{figure}

Both MWA and GMRT data sets were processed with DSPSR \citep{VanStraten2011b} and PSRCHIVE \citep{Hotan2004,VanStraten2011} to produce single-pulse archives.
A small percentage of frequency channels contaminated by local radio frequency interference (RFI) were removed, and impulsive RFI was identified by eye on a pulse-by-pulse basis; pulses that were contaminated by RFI were replaced by zeros.
The resulting Stokes I average profiles are shown in Fig. \ref{fig:obs3_psrchive}.

\begin{figure*}
    \centering
    \includegraphics[scale=0.5]{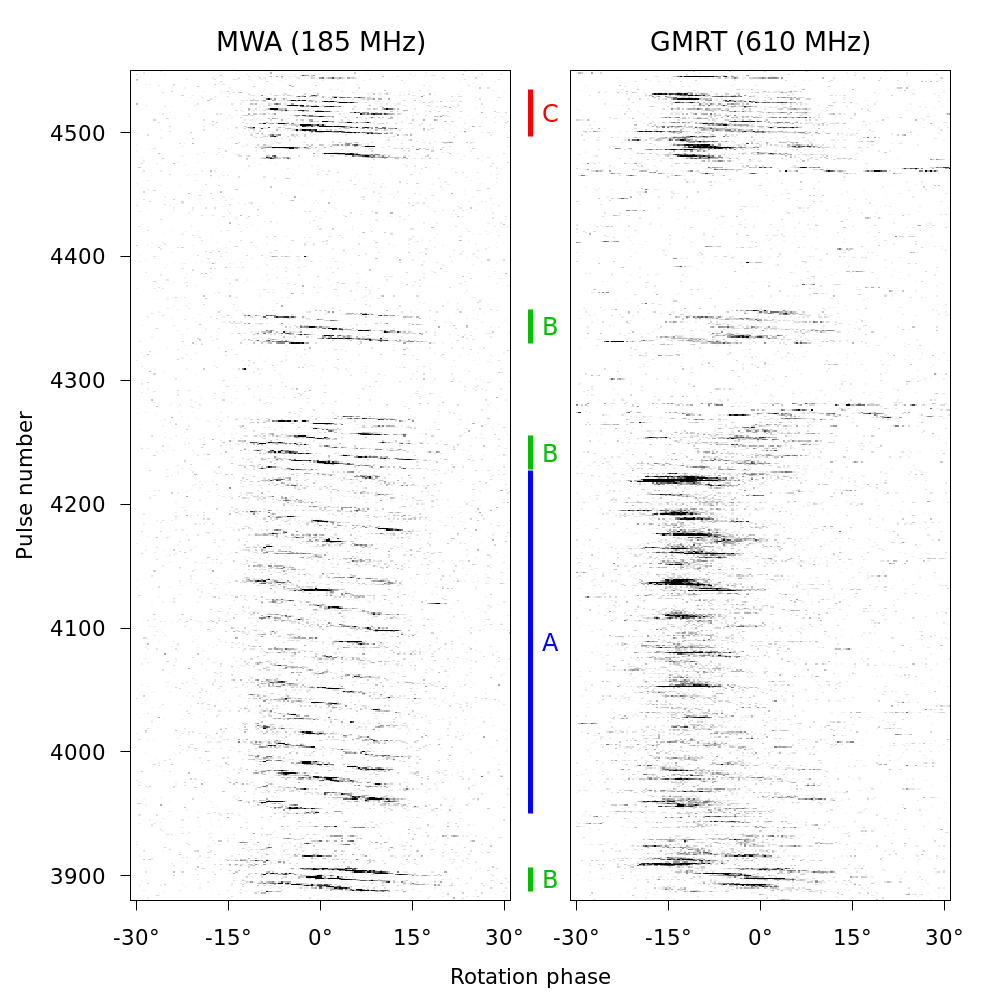}
    \caption{A sequence of 671 simultaneous pulses observed at the MWA (left) at $185\,$MHz and at the GMRT (right) at $610\,$MHz. All three drift modes can be seen (including the rare C mode), along with interspersed null sequences. The dynamic ranges have been adjusted by eye to give comparable contrast. Some pulses contaminated with RFI can be seen in the GMRT data set. As in Fig. \ref{fig:obs3_psrchive}, the central phase at $0^\circ$ was chosen by eye to approximate the centre of the pulse window.}
    \label{fig:showcase_modes}
\end{figure*}

The analysis in this paper will naturally benefit from knowledge of the exact clock offset between the two telescopes, which unfortunately was unavailable at the time of recording due to the fact that the MWA had not yet embarked on any timing programs that could verify the VCS time stamps.
The correspondence between pulses was therefore found by aligning the null sequences, which are known in the case of \psrB{} to be a broadband phenomenon \citep{Smits2007}.
The total number of pulses observed simultaneously at both telescopes was $4780$.
A section of the pulsestacks, showcasing all three drift modes, is shown in Fig. \ref{fig:showcase_modes}.

\section{Analysis \& Results}
\label{sec:anl}

\subsection{Using $P_3$ to constrain the carousel geometry}
\label{sec:drift_constraint}

In the simplest version of the carousel model, in which the surface sparks are equally spaced in magnetic azimuth, the behaviour of the drift bands is determined by the number of sparks in the carousel, $n$, and the carousel rotation rate, $P_4$.
If the carousel rotation is constant, the observed $P_3$ at any given longitude will also be constant.
This is true regardless of either the radial distance of the spark from the magnetic axis, or the sparks' radial extent.
Without aliasing, $P_3$ is identical to $P_4/n$, but with aliasing,
\begin{equation}
    \frac{1}{\overline{P}_3} = \left|\frac{n}{\overline{P}_4} - k\right|,
    \label{eqn:aliasing}
\end{equation}
where $k = [n/\overline{P}_4]$ is the aliasing order\footnote{Under this definition of the aliasing order, the sign of $k$ is always the same as the sign of $P_4$, which in this work is chosen to be positive when the carousel is rotating anticlockwise when viewed from ``above'' the magnetic axis. The phrase ``first order aliasing'' can be taken to mean $k = \pm 1$.}, and where the square brackets denote rounding to the nearest integer.
Eq. \eqref{eqn:aliasing} is the same as that derived by \citet{Rankin2013}, but generalised for arbitrary aliasing orders.

The fact that $P_3$ is different for each of \psrB{}'s three drift modes can therefore be explained by a change in $n$, $P_4$, or both.
In general, it is difficult to find the correct aliasing order $k$, because solutions for Eq. \eqref{eqn:aliasing} can always be found for any $k$ due to the free choice of $P_4$ for each mode.
However, the carousel rotational speed is thought to be set by the magnetic and electric fields near the surface, which cannot easily change their magnitudes and/or directions on the time scales on which drift mode changes are observed to take place ($< 1$ rotation).
Therefore, it is an interesting question to ask whether the behaviour of the different modes might be brought about by only a change in $n$, and keeping $P_4$ constant, an idea first introduced by \citet{Rankin2013} for PSR B1918+19.
In this case, it is immediately clear that $k$ cannot be $0$ for all three drift modes, since otherwise the drift rate would be the same at all times.
If we assume that $k$ is the same for all three drift modes, we find that the quantities $1/\overline{P}_{3A}$, $1/\overline{P}_{3B}$, and $1/\overline{P}_{3C}$ (for modes A, B, and C, respectively) are members of an arithmetic sequence whose $n$th term corresponds to a carousel with $n$ sparks.
Furthermore, if $n_A$, $n_B$, and $n_C$ are themselves an arithmetic sequence, then the respective values of $\overline{P}_{3}$ will be harmonically related, as originally suggested by \citet{Wright1981}.
It also immediately follows that
\begin{equation}
    \frac{1}{\overline{P}_{3B}} = \frac12\left(\frac{1}{\overline{P}_{3A}} + \frac{1}{\overline{P}_{3C}}\right).
\end{equation}
which can be readily verified in the case of \psrB{}.
In the special case that the number of sparks differs only by one, then the terms of the arithmetic sequence are consecutive, and the difference between any adjacent pair yields the carousel rotation rate directly, e.g.
\begin{equation}
    \left|\frac{1}{\overline{P}_{3A}} - \frac{1}{\overline{P}_{3B}}\right| = \left|\frac{1}{\overline{P}_4}\right|.
\end{equation}
Then $n$ can be found by substituting the derived value of $\overline{P}_4$ back into Eq. \eqref{eqn:aliasing}.

An alternative approach rearranges Eq. \eqref{eqn:aliasing} to leverage the expectation that the number of sparks changes by an integer value.
Continuing under the assumption that the carousel rotation does not change between drift modes, $\overline{P}_4$ can be eliminated to yield
\begin{equation}
    \frac{n_A}{\Delta n} = \frac{\overline{P}_{3B}(1 - k\overline{P}_{3A})}{\overline{P}_{3B} - \overline{P}_{3A}},
    \label{eqn:deltan}
\end{equation}
where $\Delta n = n_A - n_B$.
Using measured values for $\overline{P}_{3A}$ and $\overline{P}_{3B}$ \citep[from][summarised in Table \ref{tbl:P2P3}]{McSweeney2017}, this becomes
\begin{equation}
    \frac{n_A}{\Delta n} \approx (-1.3 \pm 0.2) + (15.9 \pm 1.7)k.
    \label{eqn:deltan_2}
\end{equation}
Plugging in candidate values for $n_A$ and $k$, we find near-integral solutions for $\Delta n$, with the first order solutions given in Table \ref{tbl:deltan}.
As expected, no solutions were found for $k=0$.
The same procedure was repeated using modes B and C in Eq. \eqref{eqn:deltan} instead of A and B, with the same results.
\begin{deluxetable}{c|cc}
    \tablecaption{Values of $P_2$ and $P_3$ for the drift modes of \psrB{}, from \citet{McSweeney2017}\label{tbl:P2P3}}
    \tablehead{\colhead{Drift mode} & \colhead{$P_2^{(\circ)}$} & \colhead{$\overline{P}_3$}}
    \startdata
    A & $18^\circ.9 \pm 1.1$ & $12.5 \pm 0.8$ \\
    B & $19^\circ.8 \pm 0.5$ &  $7.0 \pm 0.2$ \\
    C & $19^\circ.1 \pm 2.9$ &  $4.6 \pm 0.3$ \\
    \enddata
\end{deluxetable}
\begin{deluxetable}{cccccc}
    \tablecaption{Carousel parameters deduced for \psrB{}\label{tbl:deltan}}
    \tablehead{
        \colhead{$n_A$} & \colhead{$n_B$} & \colhead{$n_C$} & \colhead{$k$} & \colhead{$\sim\overline{P}_4$} & \colhead{$\Delta n$}
    }
    \startdata
    13 & 12 & 11 &  1 &  14.1 &  0.89 \\
    14 & 13 & 12 &  1 &  15.2 &  0.96 \\
    15 & 14 & 13 &  1 &  16.4 &  1.02 \\
    15 & 16 & 17 & -1 & -14.0 & -0.87 \\
    16 & 17 & 18 & -1 & -14.8 & -0.93 \\
    17 & 18 & 19 & -1 & -15.7 & -0.99 \\
    \enddata
\end{deluxetable}

Eq. \eqref{eqn:deltan_2} shows that if a change in $n$ is indeed the cause of different drift modes, then the predicted number of sparks must increase rapidly for higher aliasing orders, and more so for higher values of $\Delta n$.
If $\Delta n$ is kept constant, then $n$ increases approximately linearly with $k$, and therefore $\overline{P}_4$ must stay relatively constant (since $k = [n/\overline{P}_4]$), but will be much larger (i.e. the carousel rotates much slower) for higher values of $\Delta n$.
Thus, the approximate solution for $\overline{P}_4$ represents the fastest possible carousel rotation speeds consistent with the observed $\overline{P}_3$ values, without significant difference for higher aliasing orders.
By contrast, the theoretical value of $\overline{P}_4$ predicted by the RS model is
\begin{equation}
    \overline{P}_{4\text{,RS}} \approx 5.7 \times \left(\frac{P_1}{\text{s}}\right)^{-3/2}\left(\frac{\dot{P}}{10^{-15}}\right)^{1/2}
    \label{eqn:P4RS}
\end{equation}
which is $\overline{P}_{4\text{,RS}} \approx 4$ for \psrB{}.
Thus, we argue that a modest carousel rotation period ($14 \lesssim|\overline{P}_4| \lesssim 17$) and a modest number of sparks ($13 \le n_A \le 17$) is the most plausible, since it is much closer to the above theoretical value than a no-aliasing solution (for instance, $\overline{P}_4 = n\overline{P}_3 = 9\times12.5 = 112.5$ for mode A in \citealt{Smits2005}).
These carousel properties are similar to those found by \citet{Rankin2013} for PSR B1918+19.

On the possible harmonicity of the drift modes, we note that the finding of \citet{Vivekanand1996} that they are \emph{not} harmonically related appears to be based on the argument that the \textit{drift rates} are not harmonically related.
However, this is not the same as the claim that the values of $P_3$ are harmonically related.
Indeed, since $P_{2A} \approx P_{2B} \approx P_{2C}$, the harmonicity of $P_3$ implies that the drift rates should be linearly related, which is in fact consistent with their measurements.

\subsection{Using $P_2$ to constrain the viewing geometry}
\label{sec:P2_constraint}

The preceding section dealt solely with $P_3$, a quantity which is independent of pulse longitude, as evident in Eq. \eqref{eqn:aliasing}.
$P_2$, on the other hand, encodes information about the angular distance between successive sparks, and is generally a longitude-dependent quantity.
We now proceed to investigate how $P_2$ may be used to constrain the carousel parameters and the viewing geometry.

$P_2$ can be measured by correlation methods \citep[e.g.][]{Smits2005} or methods using fluctuation spectra \citep{Backer1971,Wolszczan1981}.
For our purposes, the latter are the most suitable for obtaining measurements of $P_2$ as a function of longitude.
This involves computing the longitude-resolved fluctuation spectrum (LRFS), which is done by applying the Discrete Fourier Transform to each longitude bin within a given drift sequence.
If the drift band separation remains constant throughout the sequence, a narrow, strong feature will appear at all longitudes in the frequency bin corresponding to the value of $P_3$.
The relative phases of the feature, which we denote by $\theta(\phase)$, constitute the so-called ``subpulse phase track'', which encodes information about $P_2$:
\begin{equation}
    \frac{1}{\overline{P}_2(\phase)} = \left|\deriv{\theta}{\phase}\right|.
\end{equation}

\begin{figure}
    \centering
    \includegraphics[scale=0.5]{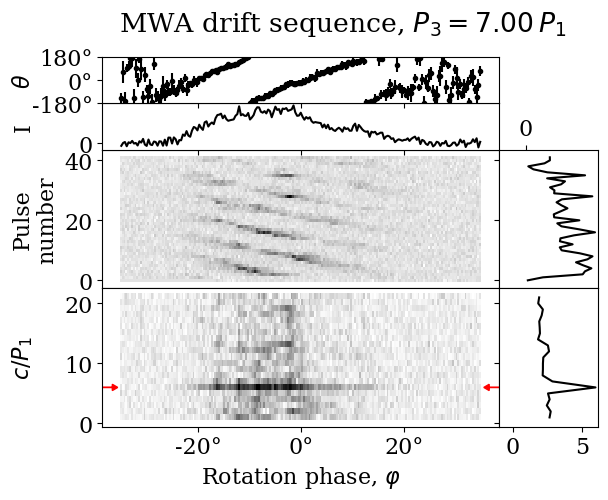}
    \includegraphics[scale=0.5]{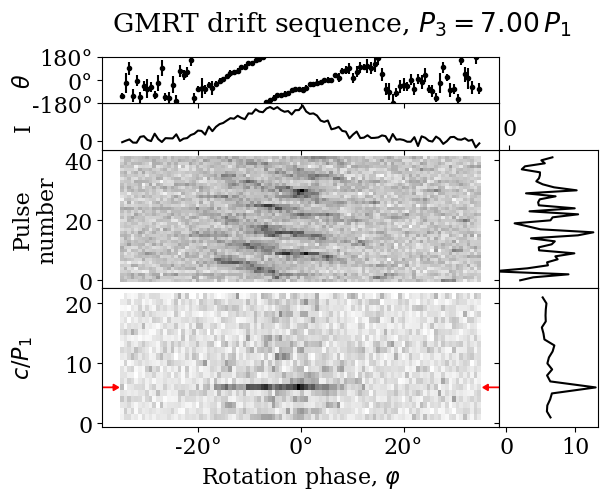}
    \caption{A mode B drift sequence observed with the MWA (top figure) and the GMRT (bottom figure). The central panel shows the pulsestack with clearly identifiable drift bands and pulse energies to the right. The LRFS is in the lower panel, with the red arrows marking the frequency bin corresponding to the vertical separation between the drift bands ($P_3$), and its integrated spectrum on the right. Immediately above the pulsestack is the integrated profile, and the subpulse phase track (i.e. the phases of the highlighted LRFS row) is at the very top. The phase centres ($\phase = 0^\circ$) have been aligned by eye.}
    \label{fig:pulsestack_example}
\end{figure}

The subpulse phase that is visible at a given rotation phase involves the computation of two quantities: (1) the magnetic azimuth, $\sigma$, of the field lines coplanar with the LoS, and (2) the spark activity at the footpoint (i.e. where the magnetic field line intersects the stellar surface).
The consideration of (1) is what \citet{Yuen2016} mean by the ``motion of the visible point.''
For a carousel of $n$ sparks rotating at rate $1/\overline{P}_4$, the subpulse phase is the difference between the sampled magnetic azimuth and the amount that the carousel has rotated in the same time:
\begin{equation}
    \theta(\phase) = n\left(\sigma(\phase) - \frac{\phase}{\overline{P}_4}\right) + \theta_0,
    \label{eqn:subpulse_phase}
\end{equation}
where the dependence on pulse phase has been made explicit, and $\theta_0$ is the subpulse phase when $\phase = 0$.
An example of a mode B sequence, observed simultaneously with the MWA and the GMRT, with their respective profiles, LRFS, pulse energies, and subpulse phase tracks, is shown in Fig. \ref{fig:pulsestack_example}.

To recover the expression for $\overline{P}_2$, we differentiate with respect to the phase, yielding\footnote{One must be careful when comparing this equation with similar-looking equations in \citet{Yuen2016}, as both the definitions and notations of the terms involved, as well as the derivation itself, differ.}
\begin{equation}
    \frac{1}{\overline{P}_2(\phase)} = n\left|\deriv{\sigma}{\phase} - \frac{1}{\overline{P}_4}\right|.
    \label{eqn:P2approx}
\end{equation}
By multiplying through by $\overline{P}_3$ and by using Eq. \eqref{eqn:aliasing} to eliminate $n$, the drift rate can be equivalently expressed as
\begin{equation}
    \frac{1}{\overline{D}} =
        \bigg(\overline{P}_3 k \pm 1\bigg)
        \bigg(\overline{P}_4 \deriv{\sigma}{\phase} - 1\bigg).
    \label{eqn:driftrate}
\end{equation}
In this form, it is easily seen that if $k = 0$, the dependence on $P_3$ vanishes and a constant carousel rotation speed implies the same drift rate for all three drift modes.
Furthermore, any observed curvature of the driftbands is necessarily inherited from the $d\sigma/d\phase$ term unless the angular speed of the sparks is a function of magnetic azimuth.

If the drift mode changes are, in fact, characterised by only a change in the number of sparks, then Eq. \eqref{eqn:P2approx} predicts only small changes in $\overline{P}_2$ for sufficiently large values of $n$.
This is because $\overline{P}_4$ is assumed constant, and both $\sigma$ and $\deriv{\sigma}{\phase}$ are pure functions of the viewing geometry.
Therefore, the fractional change in $\overline{P}_2$ should be equal to the fractional change in $n$, or
\begin{equation}
    \frac{\overline{P}_{2A}}{\overline{P}_{2B}} = \frac{n_B}{n_A},
    \label{eqn:fractionalP2}
\end{equation}
and similarly for modes A and C, etc.
The smallest integral values of $n$ that are consistent with the reported values of $P_2$ in \citet{McSweeney2017} are approximately $n \gtrsim 6$.
This is a weaker constraint than the constraint derived using $P_3$ because of the relatively large measurement errors on $P_2$ (obtained using correlation methods).

The relative magnitudes of the two terms in the brackets of Eq. \eqref{eqn:P2approx} determine the overall drift band behaviour.
For a sufficiently slowly rotating carousel, the value of $P_2$ is dominated by $d\sigma/d\phase$.
If the carousel rotates in the opposite sense to the motion of the visible point in the pulse window (i.e. $P_4$ is negative), then increasing the carousel speed causes $P_2$ to decrease.
However, if the carousel rotates in the same direction as the motion of the visible point, $P_2$ will initially increase until the two terms balance and the subpulse drift appears stationary.
As the carousel rotates faster still, the sparks overtake the visible point and the subpulse drift resumes, but now with each successive subpulse coming from the trailing spark.

Regardless of the carousel configuration, $d\sigma/d\phase$ is in general a function of rotation phase, and thus so is $P_2$.
It follows that the drift rate can never be stationary at all phases simultaneously.
In the presumably rare case that the value of $1/\overline{P}_4$ falls between the maximum and minimum value of $d\sigma/d\phase$, one would see the subpulse drift adopt opposite directions at distinct rotation phases, a phenomenon that has been observed in a few pulsars and known as ``bi-drifting'' \citep[][and references therein]{Wright2017}.
Thus, Eq. \eqref{eqn:P2approx} potentially provides a natural framework for understanding bi-drifting.

In the case of \psrB{}, we can estimate the relative contributions of the two competing terms in Eq. \eqref{eqn:P2approx} by considering what parameter values give rise to the measured $P_2$ value.
Taking, for example, the B mode of the second line in Table \ref{tbl:deltan} ($n = 13$, $\overline{P}_4 = 15.2$), a value of $d\sigma/d\phase = 1.46$ is required to reproduce $\overline{P}_2^\circ = 19.8^\circ$.
Thus, even though the carousel is rotating fast enough to produce aliasing $k = 1$, the motion of the visible point is far more dominant, since $d\sigma/d\phase \gg 1/\overline{P}_4$.
In this case, we can then write
\begin{equation}
    \deriv{\sigma}{\phase} \approx \pm \frac{1}{n\overline{P}_2},
    \label{eqn:slow_carousel_P2}
\end{equation}
which becomes an exact equality in the stationary carousel limit.
For the same viewing geometry (i.e. keeping $d\sigma/d\phase$ fixed), the contribution by the carousel rotation only becomes appreciable when either the aliasing order increases to $k \gtrsim 5$ or the number of sparks drastically increases (or some combination of both).
A similar conclusion is reached for all the entries in Table \ref{tbl:deltan}.

If we assume a small $k$ for \psrB{}, we may draw another conclusion from the observation that in the pulses immediately following the onset of a drift sequence after a null, the drift rate appears to undergo a decay until it reaches a nominally stable value \citep{McSweeney2017}.
During this decay period, both $P_2$ and the subpulse width appear visually exaggerated, suggesting that the relative speed of the visible point and the sparks is less.
However, when the motion of the visible point dominates, there are only two possibilities.
Either (1) the directional sense of both the carousel drift and the motion of the visible point is the same, in which case the relaxation of the drift rate implies a slowing down of the carousel rotation, or (2) the direction senses are different, in which case the carousel must be speeding up.
Given that the drift rate of other pulsars which are thought to be free of aliasing effects is observed to increase after a null \citep[e.g.][]{Lyne1983}, we therefore favour the second possibility.
Note that this does not necessarily imply that the carousel and the \emph{pulsar} are rotating in the same directional sense, since the sign of $d\sigma/d\phase$ depends, in part, on the sign of $\beta$, which is still unknown.

\subsection{Generalisation for AR effects}
\label{sec:ar}

The $P_2$ model derived in the previous section has no frequency dependence whatsoever.
However, $P_2$ has long been known to be a function of frequency for pulsars generally \citep{Taylor1975} and for \psrB{} in particular \citep{Bartel1980}.
More recently, \citet{McSweeney2017a} showed that the frequency-dependent shift in subpulse phase also depends on which drift mode is present.

One might expect that the dependence of $P_2$ on observing frequency is a simple side-effect of RFM, where the ``space'' between the drift bands grows in proportion to the opening angle of the emission cone \citep[e.g.][]{Smits2007,Yuen2016}.
However, it cannot be so, as discussed at length in \citet{Edwards2002,Edwards2003}.
While it is true that the line of sight would cut through the emission cone at different phases depending on the frequency, the image of a single spark projected onto the pulsar's sky would not change relative to the line of sight traverse.
Observationally, this implies that different parts of drift bands are illuminated when viewed at different frequencies, not that the drift bands themselves shift in phase.

Despite the preceding argument, a radius-to-frequency mapping remains a necessary (if not a sufficient) condition to explain the $P_2$ dependence on frequency.
For example, \citet{Bilous2018} explains the drift band shifts observed in PSR B0943+10 in terms of the offset phase of the centroid of the line of sight traverse through a beamlet of finite size and shape, as the beamlet changes its magnetic colatitude with frequency.

In this section, we consider the possibility that the frequency-dependence of the subpulse phase shift and of $P_2$ is due to aberration and retardation (AR) effects, as suggested by \citet{Edwards2003}.
If AR are the dominant effects, such a generalised model can be used to extract information about the emission heights from multi-frequency measurements of the subpulse phase.

\subsubsection{Derivation of the subpulse phase shift}
\label{sec:ar_sps}

\citet{Dyks2004} and others showed that pulsar emission will be shifted earlier in phase according to
\begin{equation}
    \phase^\prime \approx \phase - 2r^\prime,
    \label{eqn:ARshift}
\end{equation}
where $\phase^\prime$ is the observed phase, $r^\prime = r/r_L$, and $r_L = cP_1/(2\pi)$ is the light cylinder radius.
One $r^\prime$ comes from the aberration of the emission angle from the tangent to the field line, and the other comes from retardation, or the finite time it takes the light to travel from the emission point to the observer.

Although Eq. \eqref{eqn:ARshift} applies to all emission, a correct determination of the observed subpulse modulation must also account for the finite time it takes the information about the subpulse phase to travel from the surface to the emission height.
To first order, the trajectory of relativistic particles climbing from the surface to the emission point along a corotating field line is simply $r$, with a corresponding travel time equal to a phase rotation of $r^\prime$.
This means that the observed subpulse phase does not represent the spark activity at the footpoint of the field line at the time of emission, say, at phase $\phase$, but rather the spark activity at the time when the phase was $\phase - r^\prime$.
Thus, emission that comes from a given magnetic azimuth will be observed at phase $\phase^\prime \approx \phase - 2r^\prime$ and have subpulse phase
\begin{equation}
    \begin{aligned}
        \theta^\prime(\phase^\prime)
            &= n\left(\sigma(\phase) - \frac{\phase - r^\prime}{\overline{P}_4}\right) + \theta_0 \\
            &= \theta(\phase) + \frac{n}{\overline{P}_4}r^\prime.
    \end{aligned}
    \label{eqn:sp_AR}
\end{equation}

\begin{figure}
    \centering
    \includegraphics[scale=0.38]{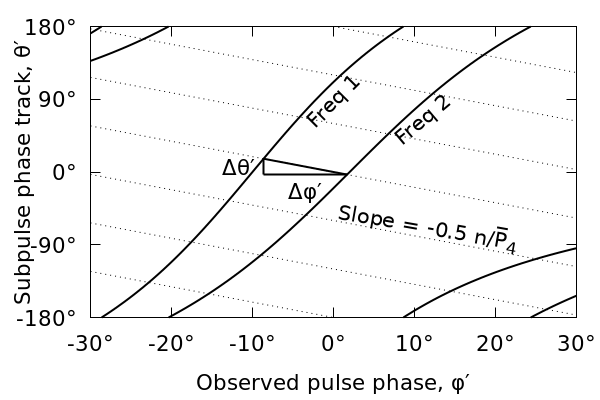}
    \caption{An illustration of the method for determining the emission height difference from the subpulse phase tracks at two different frequencies.}
    \label{fig:sp_shift_demo}
\end{figure}

Let L and H represent two field lines that have the same magnetic azimuth, $\sigma$, but whose footpoints have different magnetic colatitudes.
In the absence of AR effects, the two lines come into view at the same phase, $\phase$, at which moment the geometry is represented by the same fundamental triangle.
Let L represent the line whose footpoint has the smaller colatitude.
For a dipolar field, one expects from geometric arguments (neglecting rotation effects on curvature, as discussed in \citealt{Thomas2007} and \citealt{Thomas2010}) that the lower frequency emission will be observed from L and the higher frequency emission from H, and that the emission occurs at a greater altitude for L than for H.
From Eq. \eqref{eqn:sp_AR},
\begin{equation}
    \begin{aligned}
        \theta^\prime_\text{L}(\phase^\prime_\text{L}) - \theta^\prime_\text{H}(\phase^\prime_\text{H})
            &= \frac{n}{\overline{P}_4}(r^\prime_\text{L} - r^\prime_\text{H}) \\
            &= -\frac{n}{2\overline{P}_4}(\phase^\prime_\text{L} - \phase^\prime_\text{H}), \\
    \end{aligned}
    \label{eqn:sp_slope}
\end{equation}
where we have used Eq. \eqref{eqn:ARshift} in the second equation.
The corresponding point on the subpulse phase track will appear to have moved vertically by $\Delta\theta^\prime = \theta^\prime_\text{L}(\phase^\prime_\text{L}) - \theta^\prime_\text{H}(\phase^\prime_\text{H})$ and horizontally by $\Delta\phase^\prime = \phase^\prime_\text{L} - \phase^\prime_\text{H}$ when viewed simultaneously at two distinct frequencies.
Curiously, the slope of this apparent motion is constant:
\begin{equation}
    \frac{\Delta\theta^\prime}{\Delta\phase^\prime}
        = -\frac{n}{2\overline{P}_4}
        = -\frac12\left(k \pm \frac{1}{\overline{P}_3}\right).
\end{equation}
The shift of a whole subpulse phase track is schematically illustrated in Fig. \ref{fig:sp_shift_demo}.
If $k$ is known, the two phase tracks can be directly compared, and the phase difference gives a direct measure of the difference of the emission heights at the two frequencies:
\begin{equation}
    \Delta r^\prime \approx -\frac12 \Delta\phase^\prime,
    \label{eqn:r_from_ph}
\end{equation}
in accordance with Eq. \eqref{eqn:ARshift}.
Note that although the direction of the subpulse phase track shift is fixed by the carousel geometry, the magnitude of the shift depends on $r^\prime$, which may well be a strong function of pulse phase \citep[e.g.][]{Gangadhara2004,Thomas2010}.
This can introduce further distortions in the shape of the subpulse phase track as it appears at different frequencies.

The phase shift of individual subpulses can be estimated by Taylor expanding the rotation phase shift as a function of the corresponding subpulse phase shift and the slope of the subpulse phase track:
\begin{equation}
    \Delta\phase^\prime_\text{hor}
        \approx \Delta\phase^\prime + \Delta\theta^\prime \deriv{\phase^\prime}{\theta^\prime}.
\end{equation}
By applying Eq. \eqref{eqn:sp_slope} and identifying $\overline{P}^\prime_2 = d\phase^\prime/d\theta^\prime$ as the AR-corrected subpulse separation actually measured by the observer, we find
\begin{equation}
    \Delta\phase^\prime_\text{hor}
        \approx \Delta\phase^\prime \left(1 - \frac{n\overline{P}^\prime_2}{2\overline{P}_4}\right),
\end{equation}
which by virtue of Eq. \eqref{eqn:r_from_ph} can be rearranged to yield
\begin{equation}
    \Delta r^\prime \approx \frac{\Delta\phase^\prime_\text{hor}}{2 - k\overline{P}^\prime_2 \pm \overline{D}^\prime},
\end{equation}
which directly relates the phase shift of individual subpulses to the difference of emission heights at two simultaneously observed frequencies.

\subsubsection{Derivation of $P_2$}
\label{sec:ar_P2}

The preceding analysis can be recast explicitly in terms of $P_2$ rather than the subpulse drift phase.
It follows from Eq. \eqref{eqn:sp_AR} that the measured value of $\overline{P}_2^\prime$ (where, as before, the primes on $\overline{P}_2^\prime$ and $\phase^\prime$ denote values measured in the observer's AR-corrected frame) is
\begin{equation}
    \frac{1}{\overline{P}_2^\prime}
         = \deriv{\theta^\prime}{\phase^\prime}
        \approx \deriv{}{\phase^\prime} \left(\theta + \frac{n}{\overline{P}_4}r^\prime\right).
    \label{eqn:P2_with_AR}
\end{equation}
In order to relate this to the $\overline{P}_2$ of Eq. \eqref{eqn:P2approx}, we may apply the chain rule, noting that
\begin{equation}
    \deriv{}{\phase^\prime}
        = \deriv{\phase}{\phase^\prime}\deriv{}{\phase}
        \approx \frac{1}{1 - 2\deriv{r^\prime}{\phase}} \deriv{}{\phase},
\end{equation}
where we have taken the derivative of Eq. \eqref{eqn:ARshift} with respect to $\phase$.
Although it is possible to express these equations in terms of $dr^\prime/d\phase^\prime$, which is the rate of change of emission height with AR-corrected longitude as it would appear in the observer's frame, $dr^\prime/d\phase$ is easier to relate to the emission geometry.
Then Eq. \eqref{eqn:P2_with_AR} becomes
\begin{equation}
    \frac{1}{\overline{P}_2^\prime}
        = \frac{1}{1 - 2\deriv{r^\prime}{\phase}} \left(\frac{1}{\overline{P}_2} + \frac{n}{\overline{P}_4}\deriv{r^\prime}{\phase}\right).
\end{equation}
Two simplifying assumptions can be made.
First, as before, we expect the $n/\overline{P}_4$ term to be small compared to $1/\overline{P}_2$.
Second, $dr^\prime/d\phase$ itself must be small on average, since the maximum height is not expected to be more than a few percent of the light cylinder radius.
If $\beta$ is very small, then $r^\prime$ may change rapidly near the fiducial point, but otherwise we assume $dr^\prime/d\phase \ll 1$.
Thus, neglecting the $n/\overline{P}_4$ term and Taylor expanding the fraction,
\begin{equation}
    \frac{1}{\overline{P}_2^\prime}
        \approx \frac{1}{\overline{P}_2} \left(1 + 2\deriv{r^\prime}{\phase}\right).
    \label{eqn:P2_AR_approx}
\end{equation}
Thus, the observed subpulse separation includes both a fixed, frequency-independent term equal to the subpulse separation that would be observered in the absence of AR effects, and a frequency-dependent term that depends on the rate of change of the observed emission height with rotation phase.

In \psrB{}, $\overline{P}_2^\prime$ follows the same trend as the separation of average components, i.e. it becomes smaller at higher frequencies.
Thus, the frequency-dependent term must be positive.
Since the smallest emission height is expected to be observed at the fiducial point, this supports the idea, suggested by \citet{McSweeney2017}, that the fiducial point lies on the leading side of the pulse window.

Eq. \eqref{eqn:P2_AR_approx} is almost in a form that can be used to fit historical measurements of $P_2$ against observing frequency.
If the RFM is independent of magnetic azimuth, the emission height can be written in the form $r^\prime = \Phi(\phase) \nu^\eta$, where $\Phi(\phase)$ encapsulates the fact that the magnetic colatitudes of the footpoints of the observed field lines is a function of phase, as illustrated in the geometry of \citet{Gangadhara2001} and \citet{Gangadhara2004}.
It then follows that, for a given fixed observing frequency $\nu$,
\begin{equation}
    \deriv{r^\prime}{\phase^\prime}
        = \deriv{\Phi}{\phase^\prime} \nu^\eta
    \label{eqn:rfm_assumption}
\end{equation}
and
\begin{equation}
    \frac{1}{\overline{P}_2^\prime}
        \approx \frac{1}{\overline{P}_2} \left(1 + 2\deriv{\Phi}{\phase^\prime} \nu^\eta \right).
    \label{eqn:P2_vs_freq}
\end{equation}

\subsection{Application of the model to \psrB{}}

\begin{figure}
    \centering
    \includegraphics[scale=0.5]{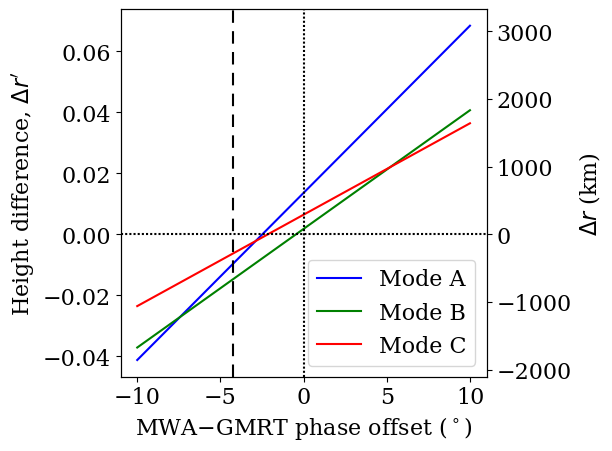}
    \caption{The relative emission height differences at MWA and GMRT frequencies for each drift mode, as a function of the assumed clock offset between the two telescopes, expressed in terms of rotation phase. A positive $\Delta r^\prime$ means that the low-frequency emission (observed at the MWA) occurs at greater altitudes than the high-frequency emission (GMRT). The zero-axes are drawn as dotted lines for reference. The $0^\circ$ phase offset corresponds to the assumption that the phase bin with the greatest average intensity at the MWA aligns with the analogous phase bin at the GMRT. The dash line at $-4.2^\circ$ is the offset assuming that a ``giant'' pulse that appears at both frequencies occurs at the same true phase.}
    \label{fig:offset_height_diff}
\end{figure}

Frequency-dependent subpulse behaviour can be thought of in terms of an absolute phase shift (a bulk shift towards earlier or later phases) and a changing $P_2$ (a compression or expansion of the subpulse modulation pattern).
However, measuring the absolute phase shift for our dual-frequency observation of \psrB{} proved problematic because the exact time offset between the clocks at the two telescopes was unknown.
The average profiles for each drift mode are known to evolve significantly with frequency, so they cannot be used to align the phases reliably\footnote{A wide-band observation using a single telescope would obviate this problem, unless the profile evolution interferes with the correct determination of DM.}.
Alternatives, such as methods involving the alignment of single pulses (e.g. via cross-correlating the time series), would mask any average subpulse phase shift that does exist between the two frequencies.
Therefore, the best that can be done with our present data sets is to show how the estimated emission height difference depends on an assumed clock offset between the two telescopes.
Using $k=1$ and the measured average values of $\overline{P}_2$ and $\overline{P}_3$ for the three drift modes, the relative emission height differences between the two frequencies are plotted for each mode in Fig. \ref{fig:offset_height_diff}.
The measured value of $\phase^\prime_\text{hor}$ was derived by cross correlating each pulse at the two frequencies, and average all the resulting correlation curves for pulses belonging to the same drift mode.
The resulting average correlation curves were smoothed with a Savitzky-Golay filter \citep{Savitzky1964} using a fifth order polynomial and then using cubic-spline interpolation to identify the location of the primary peak.

Although the MWA-GMRT clock offset is unknown, we mark two offsets of interest in Fig. \ref{fig:offset_height_diff}.
The first corresponds to assumption that the peaks of the profiles at the two respective profiles correspond to the same rotation phase.
This offset defines the zero point of the $x$-axis used in the figure.
The second offset uses the peak of a particularly bright, narrow burst that appears at both frequencies (pulse number 579 in the MWA data set, equivalent to pulse number 149 in the GMRT data set) to align the clocks.
The possibility that the two pulses occur at the same ``true'' rotation phase is founded on the identification of the bright emission as a candidate ``giant pulse'', which are known to come from very narrow phase ranges at some frequencies \citep{Kuzmin2004,Tuoheti2011,Nizamdin2011}.
If a broadband burst comes from a sufficiently small volume of the magnetosphere, as would be the case for a highly localised surface spark event that generates a large range of particle energies, then the phase shift due to AR effects would apply to all frequencies equally, and the burst would appear at a fixed longitude at all frequencies.
However, since Fig. \ref{fig:offset_height_diff} shows that this assumption leads to negative height differences in this model (the vertical dashed line at $-4.2^\circ$), we conclude that the bright emission must also come from a range of heights, thus making the derived clock offset erroneous.
We note the possibility that the GMRT bright pulse is actually a burst of RFI, as it resembles other instances of off-pulse RFI found in the data set, but we do not investigate this further.

Fig. \ref{fig:offset_height_diff} shows that the height differences for all three modes are positive when the MWA-GMRT phase offset $\gtrsim -0.^\circ5$.
In this region, the mode with the greatest $\Delta r$ is mode A, with the height differences in modes B and C being smaller but similar to each other.
For phase offsets less than $5^\circ$, $\Delta r \lesssim 2000\,$km for mode A and $1000\,$km for modes B and C (all within a few percent of the light cylinder radius).
In contrast, the height differences found by \citet{Smits2007} were $\sim 10\,$km across a much wider frequency range ($157\,$MHz to $4850\,$MHz); however, their model was based (in part) on the idea that $P_2$ is a direct manifestation of RFM, which, as we have argued above, is an incorrect interpretation.

\begin{figure}[!t]
    \centering
    \includegraphics[scale=0.3]{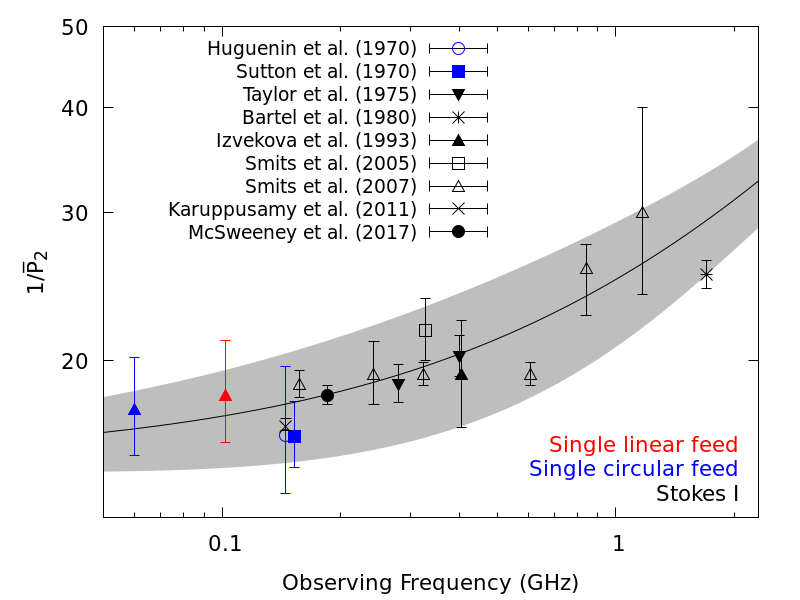}
    \caption{A re-creation of Fig. 4 of \citet{McSweeney2017}, but presented on log-scale axes, with the ordinate representing the quantity $1/\overline{P}_2$ instead of $P_2$, and including a few data points omitted previously. The errors are those reported in the respective source papers, except for \citet{Sutton1970}, which did not report measurement errors for $P_2$, and for which we have adopted a ``typical'' error of $\pm 2^\circ$. The solid line is a fit of Eq. \eqref{eqn:P2_vs_freq} to the data points. The best fit yields $\overline{P}_2^{(\circ)} = 23.^\circ6 \pm 3.^\circ8$ and $\eta = 0.70 \pm 0.34$. The gray shading shows the $\le1\sigma$ uncertainty region derived by propagating the (assumed Gaussian) errors of the fitted parameters. Omitting measurements made with single polarisation feeds still shows clear evidence that $P_2$ is a non-constant function of frequency.}
    \label{fig:P2_vs_freq}
\end{figure}

The unknown clock offset does not affect measurements of $P_2$, which is related only to the relative phase shift of subpulses at different longitudes.
Fig. \ref{fig:P2_vs_freq} shows a fit for historical measurements of $P_2$ during mode B drift sequences (mode B is the most common mode with the most reliable data, and has been observed at a much wider range of frequencies than either mode A or mode C), over a wide range of frequencies.
There is a clear general trend that $P_2$ gets smaller with increasing frequency, as previous authors have noted.
As indicated in the figure, some of the early, low-frequency measurements of $P_2$ relied on single polarisation telescope feeds.
It is unclear to what extent this would affect the measurements, given the known complexity of \psrB{}'s subpulse polarisation \citep{Taylor1975,Gould1998} and similar examples of $P_2$ ambiguity in other pulsars \citep[e.g.][]{Rankin2005}.
However, even if these particular observations were omitted, we find that there is still strong support for a frequency-dependent (i.e. non-constant) $P_2$, especially if one accepts the data from \citet{Smits2007}, whose method for measuring $P_2$ was identically used for (sometimes simultaneous) observations across a broad range of frequencies, albeit using three different telescopes.
We thus proceed on the cautious assumption that all the historical measurements presented in Fig. \ref{fig:P2_vs_freq} are valid.
Using Eq. \eqref{eqn:P2_vs_freq}, the fitted parameters are $\overline{P}_2^{(\circ)} = 23.^\circ6 \pm 3.^\circ8$, $d\Phi/d\phase^\prime = 0.3 \pm 0.1$, and $\eta = 0.70 \pm 0.34$, which is also plotted on the figure for comparison.
A discussion of this result and its implications is deferred until Section \S\ref{sec:AR_discuss}.
\nocite{Karuppusamy2011}

\section{Discussion}
\label{sec:discussion}

\subsection{Carousel geometry}

We have attempted to find plausible, quantitative explanations for two interesting features of \psrB{}'s subpulse modulation pattern: (1) the changing $P_3$ and drift rate associated with its three distinct drift modes (A, B, and C), and (2) the dependence of subpulse arrival phases on observing frequency and on drift mode.
The interpretation of both of these effects depend on the carousel model, assumed in this analysis, which connects the observed modulation pattern of subpulses to a set of sparks near the pulsar's surface rotating around the magnetic axis.
Measuring the number of sparks, $n$, and the carousel rotation period, $P_4$, is an important step in verifying the predictions of the carousel model.
However, these measurements are difficult to make, as there are often degenerate sets of parameters that produce similar looking drift bands in the pulse window.
We have argued that the most likely carousel configuration is one in which the subpulses are aliased ($k = \pm 1$), with the number of sparks being $n \approx 15$, and where $n_A$, $n_B$, and $n_C$ differ by $1$ in turn.

This solution explains a number of otherwise puzzling features of \psrB{}'s drift modes.
Firstly, the relationship between the modes' respective $P_3$ values has long been suspected to be a harmonic one.
We have shown that this is a mathematical consequence of the number of sparks differing by a constant amount (in this case, one).
Also, the similarity of $P_2$ in all three modes favours solutions with a relatively large number of sparks, as per Eq. \eqref{eqn:fractionalP2}, which is incompatible with a no-aliasing solution by virtue of Eq. \eqref{eqn:deltan_2}.
This solution has the added benefit that the different drift rates of the three modes can be understood in terms of a changing number of sparks rather than a change in the carousel drift rate itself \citep{Rankin2013}.
This is more consistent with the original claim of RS that the carousel speed is a (constant) function of the electric and magnetic field strengths at the surface, which are not expected to change rapidly on small time scales.

The absolute magnitude of the carousel rotation period is also an important prediction of the RS and competing models.
The solutions presented in Table \ref{tbl:deltan} all predict a similar carousel rotation period; taking one example ($[n_A,n_B,n_C] = [15,14,13]$), the carousel rotation period is $\overline{P}_4 = \pm 16.4$.
The time (per rotation period) it takes one spark to reach the location of its predecessor is of order $\overline{P}_4/n \approx 1$, as expected for a solution with first-order aliasing.
This is much smaller than the time that would be inferred directly from $\overline{P}_3$ if aliasing were assumed to be absent ($12.5$, $7.0$, $4.6$ for modes A, B, and C respectively).

In Table \ref{tbl:P4s}, we compare our measurement of $P_4$ (for a particular carousel configuration from Table \ref{tbl:deltan}) alongside measurements for other pulsars.
The carousel rotation period predicted from RS, $\overline{P}_{4\text{,RS}}$, is calculated from Eq. \eqref{eqn:P4RS}.
We have included both the unaliased and (first order) aliased values of $P_4$ for each pulsar in the table except B0826$-$34, whose drift bands frequently change directions, and whose unaliased $P_4$ value is therefore ambiguous.
The preferred $P_4$ (according to the cited work) is indicated in the table notes, with the other value being extrapolated assuming the same number of sparks.
We have not indicated whether $k=1$ or $k=-1$, but this can be worked out using the table values and Eq. \eqref{eqn:aliasing}.

\begin{deluxetable*}{ccCCCCl}
    \tablecaption{Carousel rotation rates of selected pulsars\label{tbl:P4s}}
    \tablehead{\colhead{Pulsar} & \colhead{Modes} & \colhead{$n$} & \multicolumn{2}{c}{$\sim|\overline{P}_4|$} & \colhead{$\overline{P}_{4\text{,RS}}$} & Reference \\
    \cline{4-5}
    & & & $k=0$ & $|k|=1$ & }
    \startdata
    \psrB{}    & [A,B,C] & [15,14,13] & [187,98,60] & 16.4 & 4.0 & [This work] \\
    B0809$+$74 & - & \gtrsim15 & \gtrsim165\tablenotemark{a} & \gtrsim14 & 1.6 & \citet{VanLeeuwen2003} \\
    B0818$-$41 & - & 20 & 370\tablenotemark{a} & 18.3\tablenotemark{a} & 1.9 & \citet{Bhattacharyya2007} \\
    B0826$-$34 & - & 13 & \text{Fluctuates in} & 13\tablenotemark{a} & 2.3 & \citet{Gupta2004,Esamdin2005}; \\
    & & & \text{both directions} & & & \quad\citet{Bhattacharyya2008} \\
    B0943$+$10 & B & 20 & 43 & 37\tablenotemark{a} & 9.3 & \citet{Deshpande2001} \\
    B1237$+$25 & - & 18\tablenotemark{b} & \sim 60 & 26\tablenotemark{a} & 3.4 & \citet{Maan2014} \\
    B1819$-$22 & [A,B,C] & [12,11,10] & [204,80,46] & 12.8\tablenotemark{a} & 2.6 & \citet{Joshi2018} \\
    B1857$-$26 & - & 20 & 147\tablenotemark{a} & 18 & 5.4 & \citet{Mitra2008} \\
    B1918$+$19 & [A,B,C] & [10,9,7] & [61,34,17] & 12\tablenotemark{a} & 7.3 & \citet{Rankin2013} \\
    \enddata
    \tablenotetext{a}{$P_4$ values in the cited texts}
    \tablenotetext{b}{Derived for the outer ring of their `A' sequence}
\end{deluxetable*}

In almost all cases, the unaliased carousel rotation period is much larger than the value predicted by RS, a point that is noted by all authors who advance the unaliased model as the preferred one.
It is commonly recognised that aliasing could be a way to account for (or at least mitigate) this discrepancy, except in the case of B0809$+$74, in which the ``memory'' of subpulse phases across null sequences was used to show that only the unaliased model is consistent with the data.
Even if aliasing were possible for this and other pulsars, there still remains a large discrepancy between the measured $\overline{P}_4$ and the RS value, most significantly for B0809$+$74, B0818$-$14, and B1237$+$25, where there is approximately an order of magnitude difference.
It is possible that in some of these cases, an even higher aliasing order may be present.

Alternative theories of spark circulation obviate the need for invoking higher order aliasing, such as the spark model of \citet{Gil2000}, which generally predicts much longer carousel rotation periods than RS.
They analyse \psrB{}, concluding that its carousel consists of five sparks in a single annulus with a radius of $s = 0.7$ of the polar cap distance, rotating at $\overline{P}_4 \approx 34$.
While this reproduces an unaliased value of $P_3$ that agrees well with the observed value for $P_3$, they do not discuss how their model may be applied to modes A and C.
However, we note that their theory predicts a maximum allowed number of sparks of $n_\text{max}=16$, which is consistent with the $k=+1$ solutions in Table \ref{tbl:deltan}.

A final point to note is that, in many cases, the first order aliasing solution predicts a value of $\overline{P}_4$ that is comparable to the number of sparks, and hence $n/|\overline{P}_4| \sim 1$.
This is not surprising, as first order aliasing implies $0.5 \le |k| = [n/|\overline{P}_4|] \le 1.5$.

\subsection{Frequency-dependent effects}
\label{sec:AR_discuss}

Armed with a plausible solution for the configuration and rotation period of the carousel, we have attempted to make sense of the frequency-dependence of its subpulse modulation pattern.
It is important to remember that the carousel model (ignoring AR and finite-spark-size effects) predicts that any given subpulse should appear at the same rotation phase at any observing frequency, regardless of the location (or movement) of their parent sparks in the polar cap, and regardless of the emission height.
Clearly, however, the subpulses of \psrB{} \emph{do} exhibit frequency- and mode-dependent phase shifts, as has been demonstrated many times in the literature, and confirmed by our own observations.

The qualitative behaviour of the subpulses can be summarised as follows.
In any given drift mode, the horizontal separation of the drift bands (i.e. $P_2$) decreases with increasing frequency \citep[however, mode A may behave more erratically at higher frequencies, see][]{Smits2007} by roughly the same amount.
There is also an absolute, frequency-dependent phase shift of the whole subpulse pattern which is different for the three modes.

\subsubsection{Absolute phase shift and implications for emission heights}

We have derived a model for the frequency-dependence of subpulse phases due to AR effects.
Regardless of whether this turns out to be the correct explanation for \psrB{}, the derived formulas quantify the subpulse phase shift that must be present due to AR effects.
Importantly, we have included in this model the time it takes information about the spark modulation to travel from the surface to the emission height, a point which (to the best of our knowledge) has not been considered in previous studies of the frequency-dependence of subpulses.
This makes the model a potentially powerful tool for testing the carousel model in cases where AR effects are indeed dominant, and where the emission heights can be measured by other (independent) means.

Our attempt to apply this model to \psrB{}, however, encountered several difficulties.
Firstly, the exact time offset between the data obtained at the two different telescope facilities is unknown, so that the absolute phase shifts of the subpulse modulation pattern in each drift mode could not be confidently measured.
This analysis is therefore only practical if the time offset is known precisely (to within a fraction of the subpulse width, $\sim$a few milliseconds for \psrB{}), or if large instantaneous bandwidths are available at a single telescope, so that the frequency evolution of the drift bands is unambiguous.
Using the relative phase shifts between the three drift modes, we have deduced that if the time offset is no more than $13\,$ms (equivalent to $5^\circ$ in pulse longitude) from the offset obtained by aligning the average profile peaks at the two frequencies, then the emission height differences between the modes is $\lesssim 2000\,$km for mode A and $\lesssim 1000\,$km for modes B and C, or approximately a few percent of the light cylinder radius.

Within this small range of time offsets, there appears to be a greater inferred height difference generally between the drift bands of mode A than for the other two modes.
If this is true, then there should also be a correspondingly greater degree of profile evolution for mode A sequences.
This is indeed observed, as can be seen in both Fig. \ref{fig:showcase_modes} and the mode-separated profiles of \citet{Smits2005}.
However, a more robust comparison requires a more precise determination of the emission heights than we have been able to accomplish here.

\subsubsection{Subpulse separation, $P_2$}

The fact that the drift bands bunch together at higher frequencies has led many authors to incorrectly associate the frequency behaviour of $P_2$ with the frequency behaviour of the average profile.
However, they must be considered independently from each other owing to the fact that any frequency dependence of the former must be explained in the context of the geometry of a single field line, whereas the latter considers the (flaring) geometry of the global magnetic field structure.
That the two effects are independent is evidenced by PSRs B0826$-$34 and B2020$+$28, whose $P_2$ appears to evolve in the opposite sense to their average profiles \citep{Bhattacharyya2008}.
In both effects, however, the RFM plays an important role, since it relates the emission frequency to the local field geometry.
We can expect, therefore, that a correct determination of the RFM, coupled with a correct understanding of which frequency-dependent effects dominate at different emission heights, would provide a quantitative explanation for both the single pulse and the average profile behaviour.

The connection between $P_2$ and the RFM is made explicit in Eq. \eqref{eqn:rfm_assumption}, which led to the prediction of how $\overline{P}_2^\prime$ changes with frequency in Eq. \eqref{eqn:P2_vs_freq}.
The functional form of \eqref{eqn:P2_vs_freq} reveals the relationship between single pulse (i.e. $P_2$) and average pulse behaviour.
The overall shape of the curve is set by the signs of of $d\Phi/d\phase^\prime$ and $\eta$, and it is instructive to consider the possibilities.
When $d\Phi/d\phase^\prime < 0$, the resulting curve has a vertical asymptote at a (maximum or minimum, depending on the sign of $\eta$) frequency of
\begin{equation}
    \nu_\text{asymptote} = \bigg(-2\deriv{\Phi}{\phase^\prime}\bigg)^{-1/\eta},
\end{equation}
at which point $P_2$ explodes.
However, no such drastic qualitative change in $P_2$ has ever been observed, so we can rule out this scenario.
When $d\Phi/d\phase^\prime > 0$, Eq. \eqref{eqn:P2_vs_freq} predicts that $P_2$ should change in the \emph{opposite} sense to the average profile components.
As noted above, this sort of behaviour has been observed (in B0826$-$34 and B2020$+$28), but it is far less common than the alternative.
It is also troubling that $d\Phi/d\phase^\prime > 0$ implies (in the geometry of \citet{Gangadhara2004}) that we are always seeing the subpulses on the trailing side of the fiducial point, which is unlikely to be true for most pulsars, especially those with large duty cycles.
Only at the fiducial point does $d\Phi/d\phase^\prime = 0$, at which point the frequency dependence vanishes altogether.

Another challenge for this model is the high-frequency prediction that $P_2$ should continue to decrease at an ever greater rate.
On this point, the erratic behaviour of $P_2$ for mode A as reported by \citet{Smits2007} has already been mentioned.
Recent work by Ilie et al. (2019, submitted) have measured a mode B value of $P_2^{(\circ)} = (20^{+7}_{-5})^\circ$ at $\sim 1.4\,$GHz, which is notably smaller than measurements at similar frequencies shown in Fig. \ref{fig:P2_vs_freq}.
Thus, $P_2$ may also behave similarly in modes A and B at high frequencies, both of which stand in stark contrast to the high-frequency prediction of the present model.

Thus, there are a number of problems with our model of how $P_2$ should change with frequency if AR are the dominant effects.
In the particular case of \psrB{}, the fit of Eq. \eqref{eqn:P2_vs_freq} to the historical measurements of $P_2$ yields positive values for both $d\Phi/d\phase^\prime$ and $\eta$.
However, a positive value for the index, $\eta$, should be viewed with strong scepticism.
All known models of RFM \citep[see, e.g.,][and references therein]{Thorsett1992} predict a negative index.
Some pulsars \emph{are} known to show unusual profile evolution, such as B1944$+$17, whose profile components move outward at higher frequencies \citep{Kloumann2010}, and others whose more central components shift in phase much less (or not at all!) than the outer components.
However, even in the unlikely scenario that these anomalies imply an unusual RFM, this cannot be the case for \psrB{}.
Low frequency observations of \psrB{} reveal that the primary profile component becomes double at $< 100\,$MHz \citep{Izvekova1993,Suleimanova2002}.
This implies (within the context of the carousel model) that the observed footpoints are indeed at smaller colatitudes at lower frequencies, i.e. the line of sight cuts the carousel more centrally at lower frequencies.
The assumed dipolar geometry of the magnetic field requires that the emission height along inner field lines is higher than for outer field lines.
We can therefore conclude that there is nothing unusual about the RFM of \psrB{} that requires exceptional consideration, and that \psrB{} is similar to most pulsars whose $P_2$ follows the same trend as the average profile components.

Given this, can we make sense of our measurement of a positive $\eta$?
One possibility is that the RFM is a strong function of magnetic azimuth (and therefore of rotation phase), rendering Eq. \eqref{eqn:rfm_assumption} invalid.
This idea has been explored by \citet{Thomas2007} and \citet{Thomas2010}, who showed how the curvature of the field lines on the leading side is augmented by rotation effects, while on the trailing side the curvature is diminished.
A more complete modelling of $P_2$ vs frequency for \psrB{} that takes rotation-induced curvature is beyond the scope of this analysis---we only note here that this effect (or some other effect that we haven't considered) may be necessary for a complete understanding of the observed frequency-dependent behaviour of $P_2$ in \psrB{} if AR effects are the dominant cause.

Alternatively, AR effects may not be the dominant frequency-dependent effects at play here, which would naturally affect our analysis of the absolute phase shifts seen in the three drift modes (Fig. \ref{fig:offset_height_diff}).
If this is the case, one alternative is the effect of the sparks having finite size and shape, as studied in B0320+39 and B0809+74 by \citet{Edwards2003} and in B0943+10 by \citet{Bilous2018}.
We do not attempt a similar analysis for \psrB{} here, but this could now be fruitfully attempted using the carousel configuration we have derived.
In general, it would be useful to compare the spark-size model with the present AR-augmented model to see which of them dominates for a given emission geometry (i.e. carousel configuration and emission height).
Alternative models, such as the fan-beam model of \citet{Dyks2015a}, may also be compared.

\section{Conclusions}
\label{sec:conclusions}

We have presented a new solution for the carousel configuration of \psrB{}, showing that the subpulse behaviour of the three drift modes can be explained with a single carousel rotation rate.
This is possible because of aliasing effects, which produce different apparent drift rates and $P_3$ values by only changing the number of sparks in the carousel.
Among the solutions consistent with the drift rates (given in Table \ref{tbl:deltan}), we give as a representative solution a carousel rotation rate of $\overline{P}_4 = 16.4$ and number of sparks in each drift mode $[n_A,n_B,n_C] = [15,14,13]$.
We have explained how first-order aliasing solutions in which the number of sparks differs between drift modes by a constant amount ($\Delta n = 1$, in this case) produce $P_3$ values that are harmonically related, which we confirm in the case of \psrB{}.
Moreover, this carousel rotation period is much closer to the theoretical value predicted by RS than estimates that assume no aliasing.

We have extended the carousel model to include AR effects and shown how they affect the pulse phase at which subpulses appear.
However, we found several problems with the model's ability to interpret the frequency dependence of $P_2$, and that these problems were not limited to \psrB{} alone, but also to the wider pulsar population in which $P_2$ usually behaves in a similar way to the average profile components (smaller separation at higher frequencies).
This indicates that either the AR-theory is incomplete (e.g. rotation induced curvature has not been included) or AR effects are not typically the dominant cause of frequency-dependent behaviour.
In either case, the inferred emission height differences for the three drift modes ($\lesssim 2000\,$km for mode A and $\lesssim 1000\,$km for modes B and C, between $185$ and $610\,$MHz), as conservative as they are, should be nevertheless regarded with caution.

Despite these difficulties, the model describes the effect that AR \emph{must} have on subpulses in the carousel model.
As such, it provides a new method for estimating emission height differences that uses the frequency evolution of the drift bands, making it independent from other methods which use the average profile.
If it can be shown that AR effects dominate the frequency-dependent behaviour of a given pulsar, this method provides a strong test of the carousel model, which uniquely assumes that the behaviour of the subpulses is governed by spark events located very close to the pulsar surface, thereby dictating the magnitude of phase shift that will be observed for a given emission height.
Therefore, applications of this AR-augmented model to other bright pulsars which exhibit subpulse drifting and whose emission heights are known or estimated by other methods will be able to conclusively determine whether the observed phase shift is consistent with the carousel model.

\acknowledgments

The authors would like to thank A. A. Deshpande for many useful and fruitful discussions, and the anonymous referee for insightful feedback that greatly improved the paper.
They acknowledge the contribution of an Australian Government Research Training Program Scholarship in supporting this research.
GW thanks the University of Manchester for Visitor status.
This scientific work makes use of the Murchison Radio-astronomy Observatory, operated by CSIRO.
We acknowledge the Wajarri Yamatji people as the traditional owners of the Observatory site.
Support for the operation of the MWA is provided by the Australian Government (NCRIS), under a contract to Curtin University administered by Astronomy Australia Limited.
The GMRT is run by the National Centre for Radio Astrophysics of the Tata Institute of Fundamental Research.
We acknowledge the Pawsey Supercomputing Centre, which is supported by the Western Australian and Australian Governments.
Parts of this research were supported by The Centre for All-sky Astrophysics is an Australian Research Council Centre of Excellence, funded by grant CE110001020.

\bibliography{biblio}

\end{document}